\documentclass[conference]{IEEEtran}

\usepackage[utf8]{inputenc} 
\usepackage[T1]{fontenc}
\usepackage{url}
\PassOptionsToPackage{hyphens}{url}
\usepackage[bookmarks=false]{hyperref}
\usepackage{cite}       
\usepackage[cmex10]{amsmath}
\usepackage{mleftright}
\mleftright
\usepackage{graphicx}         
\usepackage{booktabs}
\graphicspath{{figures/}}

\usepackage{amsthm}
\usepackage{amsmath}
\usepackage{amssymb}
\usepackage{amsfonts}

\usepackage{subcaption}
\usepackage{standalone}

\usepackage{cleveref}
\crefname{equation}{Eq}{Eqs} 
\crefrangelabelformat{equation}{(#3#1#4--#5#2#6)}

\usepackage{float}

\usepackage{tikz}
\usetikzlibrary{backgrounds, calc, chains, fit, matrix, positioning, shadows, shapes, intersections, circuits.ee}
\tikzset{input/.style={}}
\tikzset{output/.style={}}
\tikzset{op/.style={circle, draw, fill=black!10, minimum size=2.5ex, inner sep=0ex}}
\tikzset{filter/.style={rectangle, draw, thick, fill=black!10, minimum size=3.5ex, inner sep=1ex}}
\tikzset{nn/.style={trapezium, trapezium angle=80, draw, thick, fill=black!10, inner sep=1ex}}
\tikzset{branch/.style={circle, draw, thick, fill=black, minimum size=.5ex, inner sep=0ex}}
\tikzset{tensor/.style={rectangle, draw, fill=white, minimum size=2em, double copy shadow={shadow xshift=.5ex,shadow yshift=-.5ex}}}
\tikzset{rounded/.style={rounded rectangle, draw, thick, fill=black!10, minimum size=3.5ex, inner xsep=1ex}}
\tikzset{image/.style={rectangle, draw, fill=white, minimum size=2em}}
\tikzset{>=direction ee}
\tikzset{/tikz/thin/.style={line width=.9pt}}
\tikzset{/tikz/thick/.style={line width=1.4pt}}
\tikzset{every path/.style={thin}}

\usepackage{pgfplots}
\pgfplotsset{compat=1.14}
\pgfplotsset{every axis/.append style={enlargelimits={abs=3pt},grid,axis lines=left}}
\pgfplotsset{every axis plot/.append style={thick,mark size=1.5pt,line join=bevel,mark options={solid}}}
\pgfplotsset{label style={font=\small}}
\pgfplotsset{tick label style={font=\footnotesize}}
\pgfplotsset{grid style={color=black!10}}
\pgfplotsset{legend style={draw=none,opacity=.85,font=\footnotesize,cells={anchor=west,opacity=1}}}
\pgfplotsset{every non boxed x axis/.style={xtick align=center,shorten <=-.5\pgflinewidth}}
\pgfplotsset{every non boxed y axis/.style={ytick align=center,shorten <=-.5\pgflinewidth}}
\pgfplotsset{every non boxed z axis/.style={ztick align=center,shorten <=-.5\pgflinewidth}}
\pgfplotsset{/pgf/number format/1000 sep={\,}}

\usepackage{tikz}
\usetikzlibrary{arrows.meta, positioning}

\allowdisplaybreaks[2]
\begin{document}
\begin{NoHyper}

\title{Learning to Write on Dirty Paper}

\author{%
  \IEEEauthorblockN{Ezgi~{\"O}zyılkan, Oğuzhan Kubilay Ülger, Elza Erkip}
  \IEEEauthorblockA{Dept.~of Electrical and Computer Engineering \\
  New York University, New York, USA \\
  \texttt{\{ezgi.ozyilkan, kubi, elza\}@nyu.edu}} 

}

\maketitle

\begin{abstract} 
Dirty paper coding (DPC) is a classical problem in information theory that considers communication in the presence of channel state known only at the transmitter. While the theoretical impact of DPC has been substantial, practical realizations of DPC, such as Tomlinson–Harashima precoding (THP) or lattice-based schemes, often rely on specific modeling assumptions about the input, state and channel. In this work, we explore whether modern learning-based approaches can offer a complementary path forward by revisiting the DPC problem. We propose a data-driven solution in which both the encoder and decoder are parameterized by neural networks. Our proposed model operates without prior knowledge of the state (also referred to as ``interference''), channel or input statistics, and recovers nonlinear mappings that yield effective interference pre-cancellation. To the best of our knowledge, this is the first interpretable proof-of-concept demonstrating that learning-based DPC schemes can recover characteristic features of well-established solutions, such as THP and lattice-based precoding, and outperform them in several regimes.
\end{abstract}

\section{Introduction}
\label{sec:intro}

The problem of communicating over channels with known state at the transmitter is a well-studied setup in information theory~\cite{gelfand1980coding}. Costa’s seminal work~\cite{Costa}, entitled ``Writing on Dirty Paper'', showed that, for the Gaussian channel when the state, or the interference, is also Gaussian and is known noncausally at the transmitter but not at the receiver (see Fig.~\ref{fig:sys_diag}), it is possible to pre-cancel its effect entirely and achieve the same capacity as if there were no interference present. The ``dirty paper'' analogy comes from the idea that, much like writing a message on a sheet that already has dirt marks, the transmitter can incorporate knowledge of the interference (the ``dirt'') into the signal it sends, rather than attempting to erase or avoid it. While Costa’s result is theoretically elegant, practical realizations still remain an active area of research, and significant effort has been devoted to developing implementable schemes that approach the performance promised by the theory.
\let\oldthefootnote\thefootnote%
\let\thefootnote\relax\footnote{This work is supported in part by the U.S. NSF grant \#2148315.}%
\addtocounter{footnote}{-1}
\let\thefootnote\oldthefootnote

Building on this foundational insight, \textit{dirty paper coding} (DPC) has been a key \textit{nonlinear} technique in multi-user communication systems. In particular, it plays a central role in the design of broadcast strategies for multiple-input multiple-output (MIMO) downlink channels~\cite{Caire, Yu, Vishwanath, Viswanath}. In these settings, from the perspective of each user, signals intended for other users appear as interference. However, since the transmitter has noncausal knowledge of transmitted signals for all users, it can apply DPC to pre-cancel this interference. This concept has inspired a broad line of research focused on translating DPC's theoretical gains into practical multi-user systems~\cite{Sharif, Kassouf, Jiang1, Jiang2, Lin, Palomar2007MIMO, Su}, where DPC-based designs have been shown to significantly outperform their linear counterparts.

In the high SNR regime, DPC can be approximated by Tomlinson--Harashima precoding (THP)~\cite{Tomlinson, Harashima1, Wesel}, a nonlinear precoding strategy originally introduced for intersymbol interference channels. This method provides a practical means of approximating Costa’s scheme and has been adopted in multi-user communication systems~\cite{Fischer, Zu, Shenouda, Ginis} due to its relatively low computational complexity. Although it does not achieve the full capacity promised by theory~\cite{Tu, Shamai1, Urierez}, it leverages the transmitter’s knowledge of the interference and outperforms naïve strategies that treat the interference as noise.

\begin{figure}
\centering
\begin{tikzpicture}[node distance=0.55cm, auto, scale=0.5] 
  \node (w) {$V$};
  \node (encoder) [right=of w, draw, minimum width=1.2cm, minimum height=0.8cm] {Encoder}; 
  \node (X) [right=of encoder] {$X$};
  \node (plus1) [right=of X, circle, draw, inner sep=1pt] {$+$};
  \node (S) [above=of plus1] {$S$};
  \node (plus2) [right=of plus1, circle, draw, inner sep=1pt] {$+$};
  \node (N) [above=of plus2] {$N$};
  \node (decoder) [right=of plus2, draw, minimum width=1.2cm, minimum height=0.8cm] {Decoder}; 
  \node (what) [right=of decoder] {$\hat{V}$};
  \node (power) [below=0.1cm of encoder, align=center, font=\scriptsize] {$\mathbb{E}[\|X\|^2] \leq P_X$}; 

  \draw[->] (w) -- (encoder);
  \draw[->] (encoder) -- (X);
  \draw[->] (X) -- (plus1);
  \draw[->] (S) -- (plus1);
  \draw[->] (S) -- ($(S.west) + (-4.0cm, 0)$) -- ($(encoder.north) + (0.05cm, 0.01cm)$);
  \draw[->] (plus1) -- (plus2);
  \draw[->] (N) -- (plus2);
  \draw[->] (plus2) -- ($(decoder.west) + (-0.2cm, 0)$) -- (decoder); 
  \node[below=0.1cm of $(plus2.center)!.5!(decoder.west)$] {$Y$};
  \draw[->] (decoder) -- (what);
\end{tikzpicture}
\caption{\small{The dirty paper coding problem. The transmitter maps the message $V$ and known interference $S$ to an input $X$, subject to an average power constraint. The receiver observes the channel output corrupted by additive noise $N$.}}
\label{fig:sys_diag}
\end{figure}
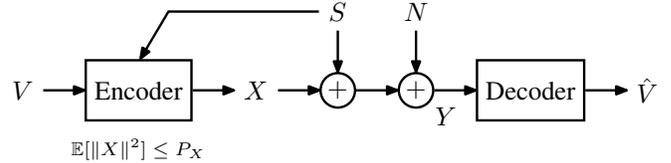

Drawing on recent advances~\cite{Whang, JSAIT2024, CISS, ISIT2023} in learning-based approaches to lossy compression with decoder-only side information (i.e., the Wyner--Ziv problem~\cite{Wyner:IT:76}), which is known to be dual of channel coding with encoder-only side information~\cite{Pradhan, Wang, Kochman}~\cite[Chapter 6]{keshet2008channel}, we formulate and demonstrate a data-driven perspective on DPC. Prior work~\cite{ISIT2023, JSAIT2024} has shown that neural networks can effectively learn nonlinear encoder–decoder mappings to perform \textit{binning}, which is a key mechanism that is known to be difficult to design analytically. Motivated by these insights, we explore how similar benefits can be realized in the DPC setting, where learning-based models are naturally suited to capture the nonlinear transforms that may aid in effective interference pre-cancellation.

As such, we revisit the classical DPC problem from a learning perspective, proposing a data-driven approach in which both the encoder and decoder are parameterized by neural networks and trained jointly. To overcome the smoothness bias that arises in artificial neural networks~\cite{rahaman2019spectral, basri2020frequency}, which makes them prone to underfitting periodic functions~\cite{ziyin2020periodic}, we incorporate theory-guided design choices by using sinusoidal activation functions. This architectural choice enables the model to recover the modulo-like behavior characteristic of THP and lattice-based precoding schemes. Furthermore, the model requires no prior knowledge of the channel, interference, or input distribution, and instead learns to adapt its behavior directly from data in the presence of known interference. It not only matches or exceeds the performance of schemes such as THP and lattice-based solutions, but also offers a learning-based reinterpretation of these well-studied designs by discovering similar underlying structures through training. To the best of our knowledge, this is the first work to demonstrate that learning-based DPC schemes can achieve such performance while remaining interpretable.

The paper is organized as follows. Sec.~\ref{sec:sys_model} reviews key theoretical results on DPC and prior constructive baselines, such as THP and lattice-based schemes. Sec.~\ref{sec:proposed_method} presents our learning-based solution. Sec.~\ref{sec:discussion} presents experimental results and discussion, followed by conclusions in Sec.~\ref{sec:conclusion}.

\section{System Model and Background}
\label{sec:sys_model}

\subsection{Gel'fand--Pinsker problem}

The discrete memoryless version of DPC problem was studied by Gel'fand and Pinsker~\cite{gelfand1980coding}, where the channel is affected by a random state (also referred to as interference) $S$ that is known noncausally at the transmitter. This channel is specified by a transition law $p(y|x,s)$, and the state $S \sim p(s)$ is independent of the input. Its capacity is given by:
\begin{equation}\label{eq:gp_capacity}
    C = \max \{ I(W;Y) - I(W;S) \},
\end{equation}
where the maximization is taken over all joint distributions of the form $p(s)p(w,x|s)p(y|x,s)$. 

For a given distribution $p(w,x | s)$, the achievability of Eq.~(\ref{eq:gp_capacity}) is established via a \textit{random binning} scheme. Specifically, $2^{nI(W;Y)}$ i.i.d. $W^n$ sequences are uniformly assigned to $2^{n(I(W;Y) - I(W;S))}$ bins. For each message, the encoder selects the corresponding bin and chooses a sequence $W^n$ that is jointly typical with the observed state sequence $S^n$. A codeword $X^n$ that is jointly typical with $(W^n, S^n)$ is then sent through the channel. Upon receiving $Y^n$, the decoder searches a unique $W^n$ among the $2^{nI(W;Y)}$ sequences that is jointly typical with the $Y^n$ index. The message is then estimated based on the bin index containing the decoded $W^n$.

\subsection{Costa's dirty paper coding result}

In this paper, we focus on the classical DPC setup shown in Fig.~\ref{fig:sys_diag}, where a transmitter wishes to communicate a message $V \in \mathcal{V}$ over an additive white Gaussian noise (AWGN) channel with independent interference $S \sim \mathcal{N}(0, \sigma_s^2)$. The received signal is given by:
\begin{equation}\label{eq:channel}
    Y = X + S + N,
\end{equation}
where $X$ is the channel input, and $N \sim \mathcal{N}(0, \sigma_n^2)$ is Gaussian noise independent of both $X$ and $S$. The transmitter is subject to an average power constraint $\mathbb{E}[\|X\|^2] \leq P_X$.
The interference $S$ is known noncausally at the transmitter but remains unknown to the receiver.

Costa’s seminal work~\cite{Costa} showed that interference known noncausally at the transmitter, as modeled in Eq.~(\ref{eq:channel}), does not reduce capacity. Under this setting, the capacity of the dirty paper channel is given by
$C_{DPC} = \tfrac{1}{2}\log\bigl( 1 + \tfrac{P_X}{\sigma_n^2}\bigr)$, which matches the capacity of a standard AWGN channel without interference. The achievability of this result is analogous to Gel'fand and Pinsker's construction and relies on typicality and random binning arguments for a Gaussian auxiliary variable $W$ (see Eq.~(\ref{eq:gp_capacity})). Extensions~\cite{Urierez, lapidoth} further show that the same capacity can be achieved for arbitrary interference distributions, provided the noise remains Gaussian.

\subsection{Tomlinson--Harashima precoding (THP) and lattice-based schemes}
\label{sec:THP_and_lattice}

A practical approximation of DPC can be achieved using THP, which operates on a symbol-by-symbol basis using scalar modulo operations. These scalar modulo operations effectively implement a structured form of binning, replacing the random binning used in the original DPC achievability scheme. The resulting many-to-one mapping enables cancellation of interference $S$ without incurring excessive power. Originally developed for channels with intersymbol interference, THP has been widely adopted in DPC scenarios due to its simplicity~\cite{Windpassinger}.

For a given message $v \in \mathcal{V} \subseteq R$, the channel input is constructed as follows:
\begin{equation}\label{eq:thp_enc}
    X = v - S -U \hspace{-.5em} \mod \Delta , 
\end{equation}
where $\Delta \in \mathbb{R^+}$ is the modulo base, and $U \sim \mathrm{Unif}(-\Delta/2,\Delta/2)$ is a shared dither known to both transmitter and receiver. The addition of this dither ensures that $X$ is also uniformly distributed over $(-\Delta/2,\Delta/2)$, which helps satisfy the power constraint. 
At the receiver, the dither is added back, and the same modulo operation is applied:
\begin{equation}\label{eq:eff_channel}
    \tilde{Y} = Y + U \hspace{-.5em} \mod \Delta = v + N \hspace{-.5em} \mod \Delta, 
\end{equation}
using Eqs.~(\ref{eq:channel}) and (\ref{eq:thp_enc}). This allows THP to completely cancel the interference $S$. For large $\Delta$ (corresponding to high SNR), the resulting channel approximates an AWGN channel. Although THP is computationally efficient and effective at high SNR, its performance degrades at low SNR due to the combined effects of dithering and the modulo operation.

THP can be interpreted as a special case of lattice-based DPC strategies \cite{nested_Zamir, Urierez}, where a higher dimensional modulo-lattice operation replaces the one-dimensional scalar modulo operation used in THP (as in Eqs.~\eqref{eq:thp_enc} and~\eqref{eq:eff_channel}). In these lattice-based schemes, the message $v \in \mathcal{V} \subseteq \mathbb{R}^n$ is precoded using a structured lattice $\Lambda$, resulting in the transmitted signal:
\begin{equation}
    X = v - \alpha S - U \hspace{-.5em} \mod \Lambda, \label{eq:lattice_1}
\end{equation}
where $\alpha \in \mathbb{R}$ is a scaling factor, and $U$ is a random dither vector uniformly distributed over the fundamental Voronoi region of the lattice~\cite{zamir}. The modulo operation in this context corresponds to subtracting the nearest lattice point:
\begin{equation}
    z \hspace{-.5em} \mod \Lambda = z - Q_{\Lambda}(z),
    \label{eq:lattice_2}
\end{equation}
where $Q_{\Lambda}(z)$ denotes the nearest neighbor quantizer with respect to the lattice $\Lambda$. This operation maps the signal into the fundamental region of the lattice, and the use of uniform dither ensures that the channel input $X$ is uniform over this region. Next, the receiver computes:
\begin{equation}
    \Tilde{Y} = \alpha Y + U \hspace{-.5em} \mod \Lambda. \label{eq:lattice_3}
\end{equation}
As shown in~\cite{Urierez}, the resultant equivalent channel can be expressed as an additive noise channel given by:
\begin{equation}
    \Tilde{Y} = v + N' \hspace{-.5em} \mod \Lambda , \label{eq:eq_channel}
\end{equation}
where $N' = (1- \alpha)U + \alpha N \mod \Lambda$. 

Since the encoder output is uniformly distributed over the fundamental Voronoi region of the lattice, its power and entropy are directly determined by the shape of the chosen lattice. For a fixed volume (i.e., fixed entropy), certain lattices can achieve lower average power than the hypercube, particularly in high dimensions~\cite{zamir}. This power gain translates into a corresponding increase in mutual information when modulo-lattice coding is applied for DPC~\cite{Urierez}.

\subsection{Modeling assumptions}
\label{subsec: performance}

In this paper, we consider an \textit{uncoded} transmission setting, where the encoder and decoder can be viewed as interfacing with conventional channel coding blocks. This setup isolates the impact of learning on the encoder–decoder mappings and highlights the ability of neural networks to model nonlinear transforms useful for interference pre-cancellation. The input to the encoder consists of uniformly distributed messages $v \in \mathcal{V}$. We assume that these messages are initially mapped to a one or two-dimensional modulation scheme, such as BPSK or QPSK, which are subsequently encoded based on interference. Our approach focuses on learning the encoder and decoder mappings to minimize the symbol error rate (SER).

\section{Neural Dirty Paper Coding Scheme}
\label{sec:proposed_method}

We consider a learning-based one-shot DPC scheme in which both the encoder and decoder, depicted in Fig.~\ref{fig:sys_diag}, are parameterized by neural networks and trained end-to-end. The encoder is represented by a deterministic function $e_{\theta}: \mathcal{V} \times \mathbb{R}^k \rightarrow \mathbb{R}^{k}$, where $\theta$ denotes its parameters, $V \in \mathcal{V}$ is the message index to be transmitted, and $S \in \mathbb{R}^k$ is the interference known noncausally by the transmitter. The state and output dimension $k \in \{1, 2\}$ are determined by the initial modulation scheme, e.g., $k = 1$ for BPSK and $k = 2$ for QPSK. The encoder outputs the channel input $X \triangleq e_{\theta}(V, S)$, which is subject to an average power constraint.  

The decoder is modeled as a probabilistic function~$p_{\phi} : \mathbb{R}^k \rightarrow \mathcal{P}(\mathcal{V})$, where $\phi$ denotes its parameters, and $\mathcal{P}(\mathcal{V})$ is the probability simplex over the message set. Given the received channel output $Y = y \in \mathbb{R}^k$, the decoder outputs a probability distribution over possible messages. During inference, we assume the decoder makes hard decisions, and the final message estimate is obtained by selecting the most likely message:
\begin{equation}
    \hat{v} \triangleq \arg\max_{v \in \{1, \cdots , |\mathcal{V}|\}} p_{\phi}(v \vert y), \label{eq:hard_dec}
\end{equation}
which results in $\mathrm{SER} = P(V \neq \hat{V})$. The entire neural DPC scheme is trained end-to-end by minimizing the objective:
\begin{equation}
L(\theta, \phi) = \mathbb{E}\left[-\log p_{\phi}(V \vert Y) + \lambda \|e_{\theta}(V, S)\|^2 \right], \label{eq:loss_fn}
\end{equation}
where the first term is the cross-entropy of the predicted distribution $p_{\phi}(v \vert y)$ relative to the true conditional distribution of the message index $V$ given the channel output $Y$, serving as a surrogate for minimizing SER. The second term penalizes the $\ell_2$-norm of the encoder output, $X = e_{\theta}(V, S)$, enforcing an average power constraint. $\lambda$ controls the trade-off between power efficiency of the encoder mapping and the accuracy of symbol detection.

\begin{figure}
    \begin{subfigure}{0.493\columnwidth}
        \centering
    \includegraphics[width=\textwidth]{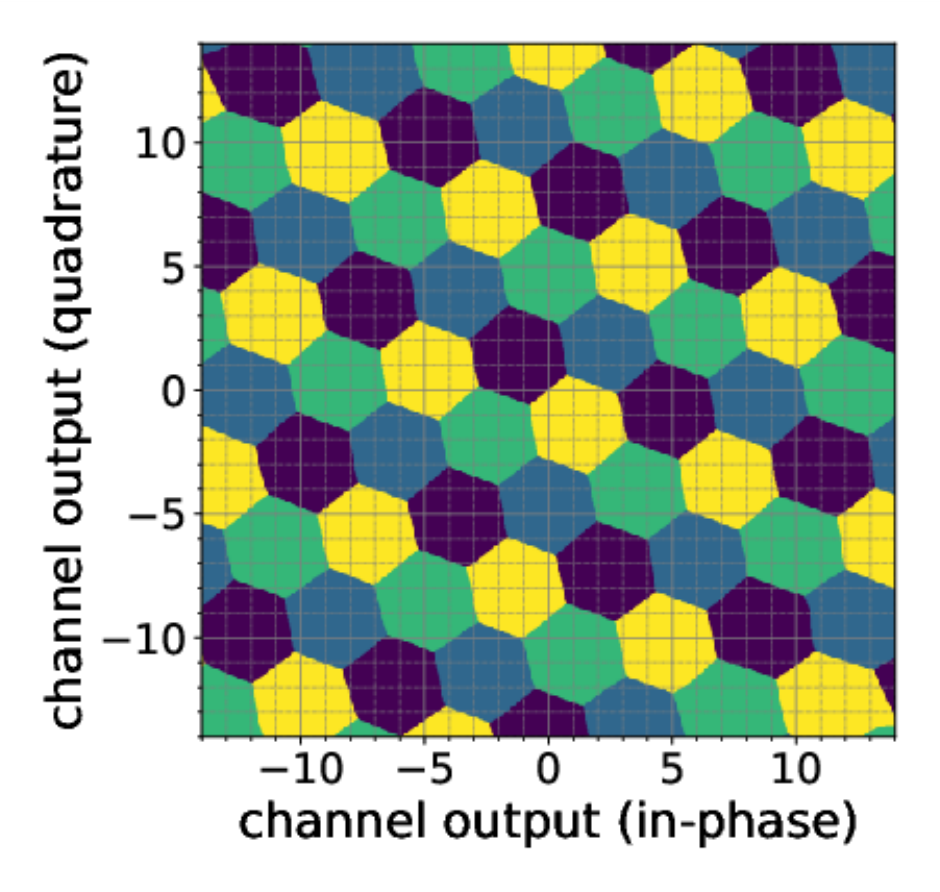}\caption{\footnotesize{w/ sinusoidal activations, scoring $\mathrm{SNR}:$ 7.03 dB,  $\mathrm{SER}:$ -1.10 dB.}}
        \label{fig:decoder_viz_sin}
    \end{subfigure}
    \hfill 
    \begin{subfigure}{0.493\columnwidth}
        \centering
        \includegraphics[width=\textwidth]{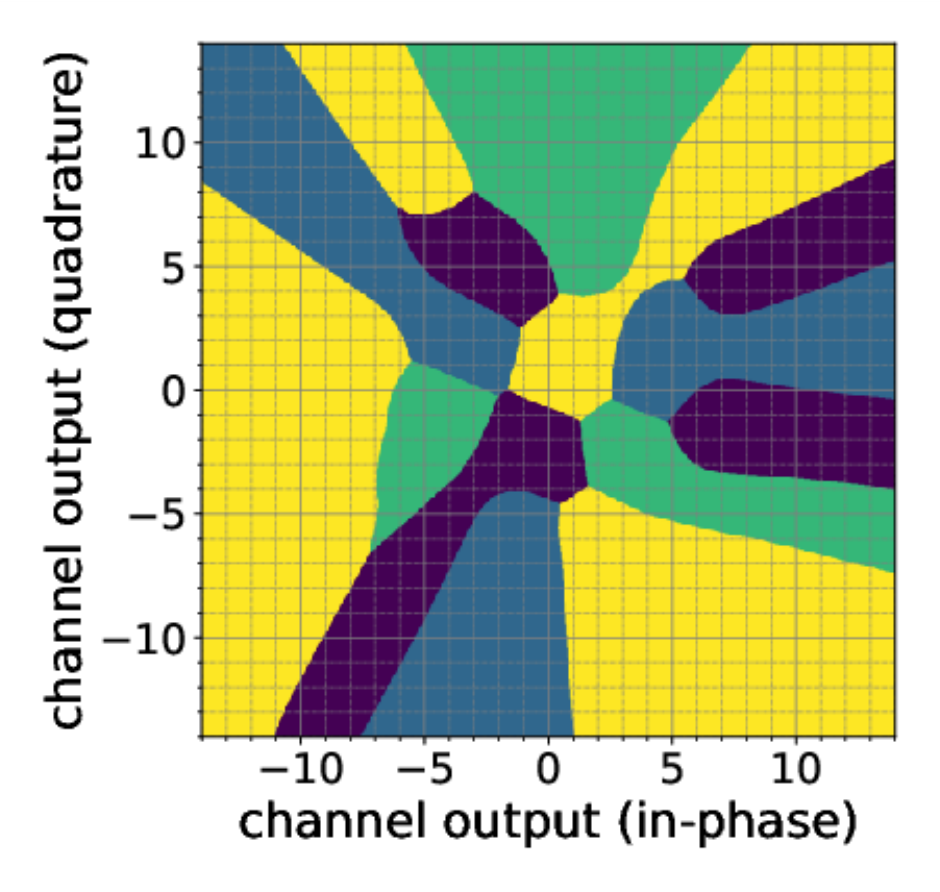}
    \caption{\footnotesize{w/ leaky ReLU activations, scoring $\mathrm{SNR}:$ 8.83 dB,  $\mathrm{SER}: $ -1.11 dB.}}
        \label{fig:decoder_viz_relu}
    \end{subfigure}
    \caption{\footnotesize{Visualization of decision regions for the proposed learning-based decoder, where the message $V$ is initially mapped to QPSK, with interference $S \sim \mathcal{N}(0, 30)$ and channel noise $N \sim \mathcal{N}(0, 1)$. The models in both panels are trained with $\lambda=5$ in Eq.~\eqref{eq:loss_fn}. The left figure shows the case where both the encoder and decoder use sinusoidal activation functions, resulting in a highly regular tiling reminiscent of the hexagonal lattice, which has the tightest sphere packing in two dimensions~\cite[Chapter 3]{zamir}. In contrast, the right figure shows the decision map when leaky rectified linear unit (ReLU) activations are used instead (with all other parameters unchanged), leading to irregular and less structured decision regions.}}
    \label{fig:decoder_mappings}
\end{figure}

Both the encoder and decoder are implemented as neural networks with three fully connected layers of 128 units each, excluding the output layers. The encoder takes as input the concatenation of the message index $v$ and the interference $s$, and maps it to a channel input $x \in \mathbb{R}^k$. All hidden layers use sinusoidal activations, which we found particularly effective for learning many-to-one mappings for the DPC setup under consideration. Consistent with the findings in~\cite{ziyin2020periodic}, we hypothesize that the sinusoidal activations enable the network to recover periodic mappings, which are essential for capturing the modulo-like behavior inherent in structured precoding schemes such as THP and lattice-based methods, as discussed in Sec.~\ref{sec:THP_and_lattice}. We observed that sinusoidal activations naturally support learning mappings that repeat over the interference domain, much like the tiling behavior of a quantizer or modulo-lattice operation as in Eqs.~\eqref{eq:lattice_1}--\eqref{eq:lattice_3}. As shown in both panels of Fig.~\ref{fig:decoder_mappings} (which will be further discussed in Sec.~\ref{sec:discussion}), replacing sinusoidal activations (Fig.~\ref{fig:decoder_viz_sin}) with standard leaky rectified linear units (ReLU) (Fig.~\ref{fig:decoder_viz_relu}) significantly alters the mapping recovered at the decoder, disrupting the emergence of a hexagonal tessellation in this case.

\begin{figure*}[t]
\begin{subfigure}{\columnwidth}
    \raggedleft
 \begin{tikzpicture}
\begin{axis}[
	scale only axis,
	font=\normalsize,
	width=\linewidth,
	height=0.65\linewidth,
	xmajorgrids=true,
	ymajorgrids=true,
	xmin=-2,
	xmax=8,
        ymin=-2.1,
	ymax=-0.2,
        xtick={-2,0,...,8},  
        ytick={-2.00, -1.75, ..., -0.25},
        y tick label style={
        /pgf/number format/.cd,
            fixed,
            fixed zerofill,
            precision=2,
        /tikz/.cd
        },
	xlabel={$\mathrm{SNR}$ $(P_X/\sigma_n^2)$ [dB]},
    ylabel={$\mathrm{SER}$ [dB]},
	legend columns=1,
	legend cell align={left},
	legend style={at={(0,0)}, anchor = south west}
	]

\addplot[line width=2pt, brown, mark=square*] table[x=X, y=Y, col sep=tab] 
    {figures/bounds/BPSK_THP.txt};
    
\addplot[line width=2pt, red, mark=diamond*] table[x=X, y=Y, col sep=tab] 
     {figures/data/BPSK_Ours_v9.txt};

\addplot[line width=2pt, green, mark=triangle*] table[x=X, y=Y, col sep=tab] 
     {figures/data/BPSK_Ours_v9_tested_var_1.0.tex};

\addplot[line width=2pt, blue, mark=pentagon*] table[x=X, y=Y, col sep=tab] 
     {figures/data/BPSK_Ours_v9_tested_var_0.5.tex};

\addplot[line width=2pt, orange, mark=o] table[x=X, y=Y, col sep=tab] 
     {figures/data/BPSK_Ours_v9_tested_var_0.1.tex};

\addplot[line width=2pt, black, mark=none, dashed] table[x=X, y=Y, col sep=tab] 
    {figures/bounds/BPSK_Lower.txt};

\legend{
    THP w/ BPSK input + $\mathbb{Z}^{1}$,
    neural DPC ($\sigma_{s, \mathrm{test}}^ 2 = 30.0$),
    neural DPC ($\sigma_{s, \mathrm{test}}^ 2= 1.0$),
    neural DPC ($\sigma_{s, \mathrm{test}}^ 2= 0.5$),
    neural DPC ($\sigma_{s, \mathrm{test}}^ 2= 0.1$),
    BPSK lower bound (AWGN),
}
     
\end{axis}

\end{tikzpicture}
  \caption{$V$ mapped to BPSK: $\mathcal{V} = \{\pm 1\}$}
  \label{fig:plot_bpsk}
\end{subfigure}%
\hfill%
\begin{subfigure}{\columnwidth}
    \raggedleft
 \begin{tikzpicture}
\begin{axis}[
	scale only axis,
	font=\normalsize, 
	width=\linewidth,
	height=0.65\linewidth,
	xmajorgrids=true,
	ymajorgrids=true,
	xmin=-2,
	xmax=8,
        xtick={-2,0,...,8},
	ymin=-1.85,
	ymax=0.05,
        ytick={-1.75, -1.5, ..., 0},
        y tick label style={
        /pgf/number format/.cd,
            fixed,
            fixed zerofill,
            precision=2,
        /tikz/.cd
        },
	xlabel={$\mathrm{SNR}$ $(P_X/\sigma_n^2)$ [dB]},
        ylabel={$\mathrm{SER}$ [dB]},
	legend columns=1,
	legend cell align={left},
	legend style={at={(0,0)}, anchor = south west}
	]

\addplot[line width=2pt, pink, mark=halfcircle*] table[x=X, y=Y, col sep=tab] 
    {figures/bounds/QPSK_ConstructionA.txt};
    
\addplot[line width=2pt, brown, mark=square*] table[x=X, y=Y, col sep=tab] 
    {figures/bounds/QPSK_THP.txt};
    
\addplot[line width=2pt, red, mark=diamond*] table[x=X, y=Y, col sep=tab] 
    {figures/data/QPSK_Ours_v8.txt};

\addplot[line width=2pt, green, mark=triangle*] table[x=X, y=Y, col sep=tab] 
    {figures/data/QPSK_Ours_v8_tested_var_1.0.tex};

\addplot[line width=2pt, blue, mark=pentagon*] table[x=X, y=Y, col sep=tab] 
    {figures/data/QPSK_Ours_v8_tested_var_0.5.tex};

\addplot[line width=2pt, orange, mark=o] table[x=X, y=Y, col sep=tab] 
    {figures/data/QPSK_Ours_v8_tested_var_0.1.tex};

\addplot[line width=2pt, mark=none, black, dashed] table[x=X, y=Y, col sep=tab] 
    {figures/bounds/QPSK_Lower.txt};

\legend{
    THP w/ QPSK input + \textit{Construction A},
    THP w/ QPSK input + $\mathbb{Z}^{2}$,
    neural DPC ($\sigma_{s, \mathrm{test}}^ 2 = 30.0$),
    neural DPC ($\sigma_{s, \mathrm{test}}^ 2 = 1.0$), 
    neural DPC ($\sigma_{s, \mathrm{test}}^ 2 = 0.5$), 
    neural DPC ($\sigma_{s, \mathrm{test}}^ 2 = 0.1$), 
    QPSK lower bound (AWGN),    
}
     
\end{axis}

\end{tikzpicture}
  \caption{$V$ mapped to QPSK: $\mathcal{V} = \{ (1 / \sqrt{2})(\pm 1 \pm i) \}$}
    \label{fig:plot_qpsk}
\end{subfigure}%
\caption{\footnotesize{Symbol error rate (SER) as a function of SNR, where the message $V$ is initially mapped to a BPSK constellation (left) or QPSK constellation (right), and the interference $S$ follows a Gaussian distribution. During training, the interference is drawn as $S \sim \mathcal{N}(0, \sigma^2_{s, \mathrm{train}}=30)$, and the channel noise is $N \sim \mathcal{N}(0, 1)$. At test time, the same trained models are evaluated under varying interference power levels, with $\sigma_{s, \mathrm{test}}^2 \in \{30.0, 1.0, 0.5, 0.1\}$. Each marker on the neural DPC curves corresponds to a separate model trained with a different value of $\lambda$ in the objective function defined in Eq.~\eqref{eq:loss_fn}.}}
\label{fig:empirical_results}
\end{figure*}
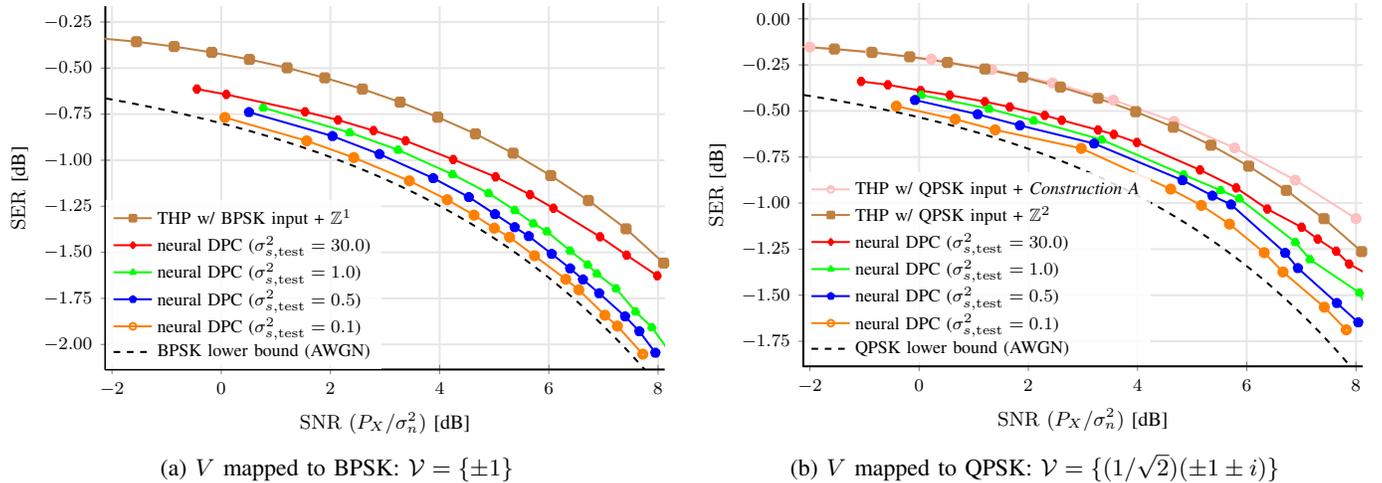

We implement our experiments\footnote{Our code is available at \texttt{\url{https://github.com/kubiulger/learned\_dirty\_paper\_coding\_ITW2025}}.} in the JAX framework~\cite{jax}, and train the neural DPC schemes using the Adam optimizer~\cite{adam} for 500 epochs, by which point the loss is visibly converged. All empirical evaluations, including power and SER estimates, are obtained by averaging over $2^{20}$ samples.

\section{Results and Discussion}
\label{sec:discussion}

For the channel model in Eq.~\eqref{eq:channel}, the interference $S$ may follow one of two models: a Gaussian distribution or a discrete distribution over a fixed modulation constellation. The former corresponds to the setting studied in Costa’s seminal paper~\cite{Costa}, where the interference is modeled as Gaussian noise independent of the message. The latter captures structured interference patterns that typically arise in multi-user communication settings, such as inter-user interference from simultaneously transmitted symbols of a downlink channel. The signal-to-noise ratio is defined as $\mathrm{SNR} = P_X / \sigma_n^2$.

For the case where $S$ follows a Gaussian distribution, we consider two standard modulation formats for the initial mapping of the messages $V$: BPSK with $\mathcal{V} = \{\pm 1\}$ and QPSK with $\mathcal{V} = \bigl\{ 1/\sqrt{2}(\pm 1 \pm i) \bigr\}$. Since the encoder is parametrized by a neural network (Sec.~\ref{sec:proposed_method}), the specific constellation points are not critical, as they only determine the cardinality of the message set $\mathcal{V}$. For the \textit{structured interference} setting, we assume that $S$ is drawn uniformly from a QPSK alphabet $\mathcal{S} = \bigl\{\sqrt{P_S/2}(\pm 1 \pm i)\bigr\}$, where $P_S$ denotes the average power of the interference.
At test time, we evaluate the performance of all learned DPC schemes in terms of the trade-off between SNR and SER. The SNR is computed by measuring the average transmit power $P_{X}$ induced by the encoder for a given value of the trade-off parameter $\lambda$ in Eq.~\eqref{eq:loss_fn}, while the SER is reported as $\mathrm{SER} = P(V \neq \hat{V})$, following Eq.~\eqref{eq:hard_dec}. 

\subsection{Baselines}
To evaluate the performance of our learned DPC schemes, we compare them against several standard baselines that reflect both theoretical limits and established practical precoding strategies (see Sec.~\ref{sec:THP_and_lattice}). The AWGN lower bounds for BPSK and QPSK correspond to the SER achieved over a standard AWGN channel without interference.

As practical benchmarks, we include THP and lattice-based schemes. A simple yet illustrative family of lattices we consider is the class of cubic lattices. In one dimension, this corresponds to the set of integers $\mathbb{Z}$, or any scaled version thereof. For BPSK input, we use a one-dimensional THP scheme based on the lattice $\Lambda = \mathbb{Z}^1$, corresponding to scalar modulo operations, as described in Eqs.~\eqref{eq:lattice_1}--\eqref{eq:lattice_3}. For QPSK input, we explore two variants of two-dimensional THP. The first applies independent scalar modulo operations to the in-phase and quadrature components, corresponding to the rectangular lattice $\Lambda = \mathbb{Z}^2$. The second method uses a 
standard approach for constructing lattices from linear codes, known as \textit{Construction A}~\cite{conway1988sphere}, which yields periodic tessellations in the cubic lattice $\mathbb{Z}^n$. Specifically, we adopt the two-dimensional lattice from the example in~\cite[Sec.~IV]{Urierez}, excluding channel coding, and select four points in its fundamental Voronoi region to define the message set $\mathcal{V}.$ We evaluate the performance of all these setups based on the equivalent channel in Eq.~\eqref{eq:eq_channel}.

\begin{figure}
    \centering
    \begin{subfigure}{0.9\columnwidth}
        \centering
        \includegraphics[width=\textwidth]{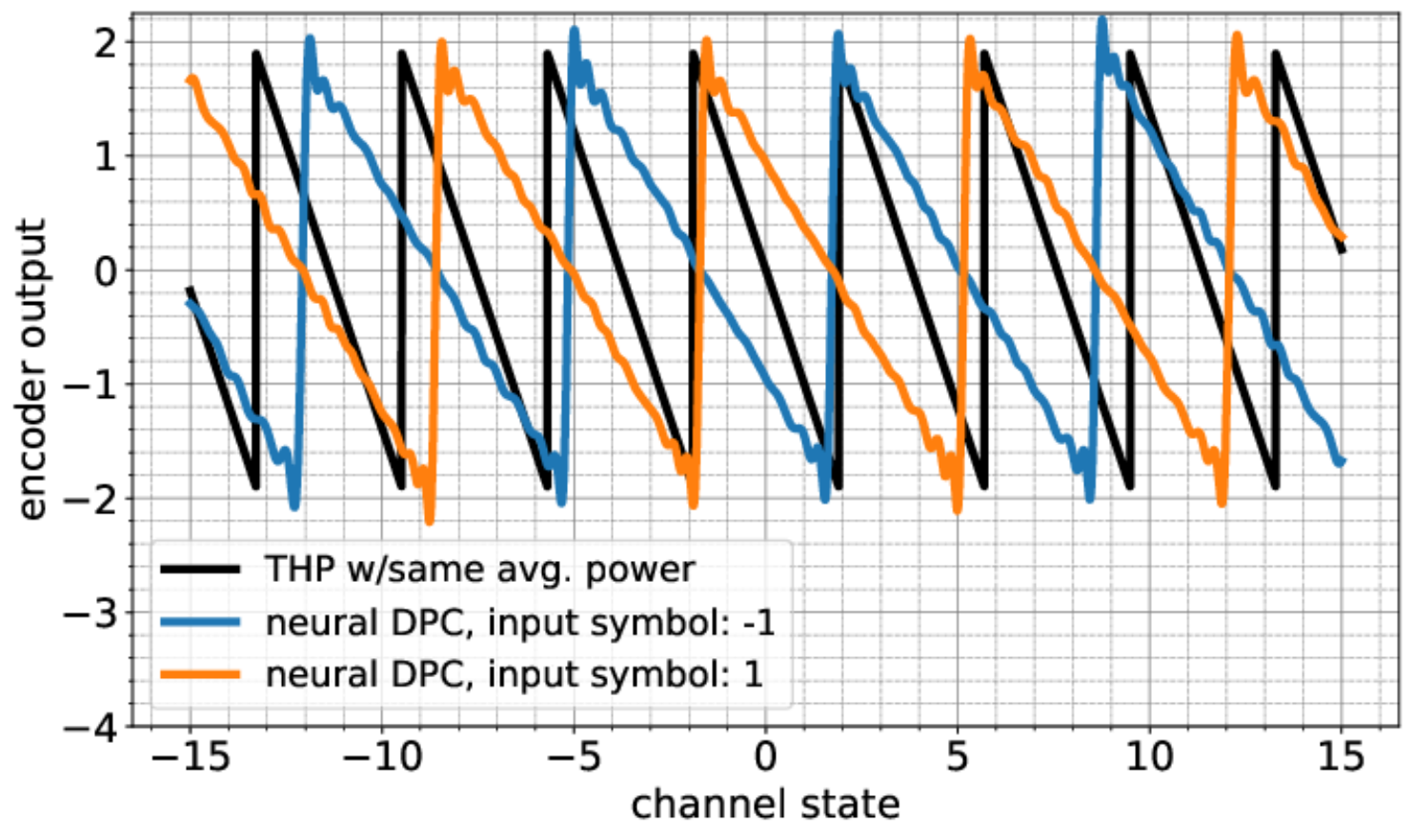}
        \caption{$\lambda=4.0$, scoring $\mathrm{SNR}:$ 0.80 dB,  $\mathrm{SER}:$ -0.68 dB.}
        \label{fig:bpsk_viz_small_snr}
    \end{subfigure}
    \vspace{1mm} 
    \begin{subfigure}{0.9\columnwidth}
        \centering
        \includegraphics[width=\textwidth]{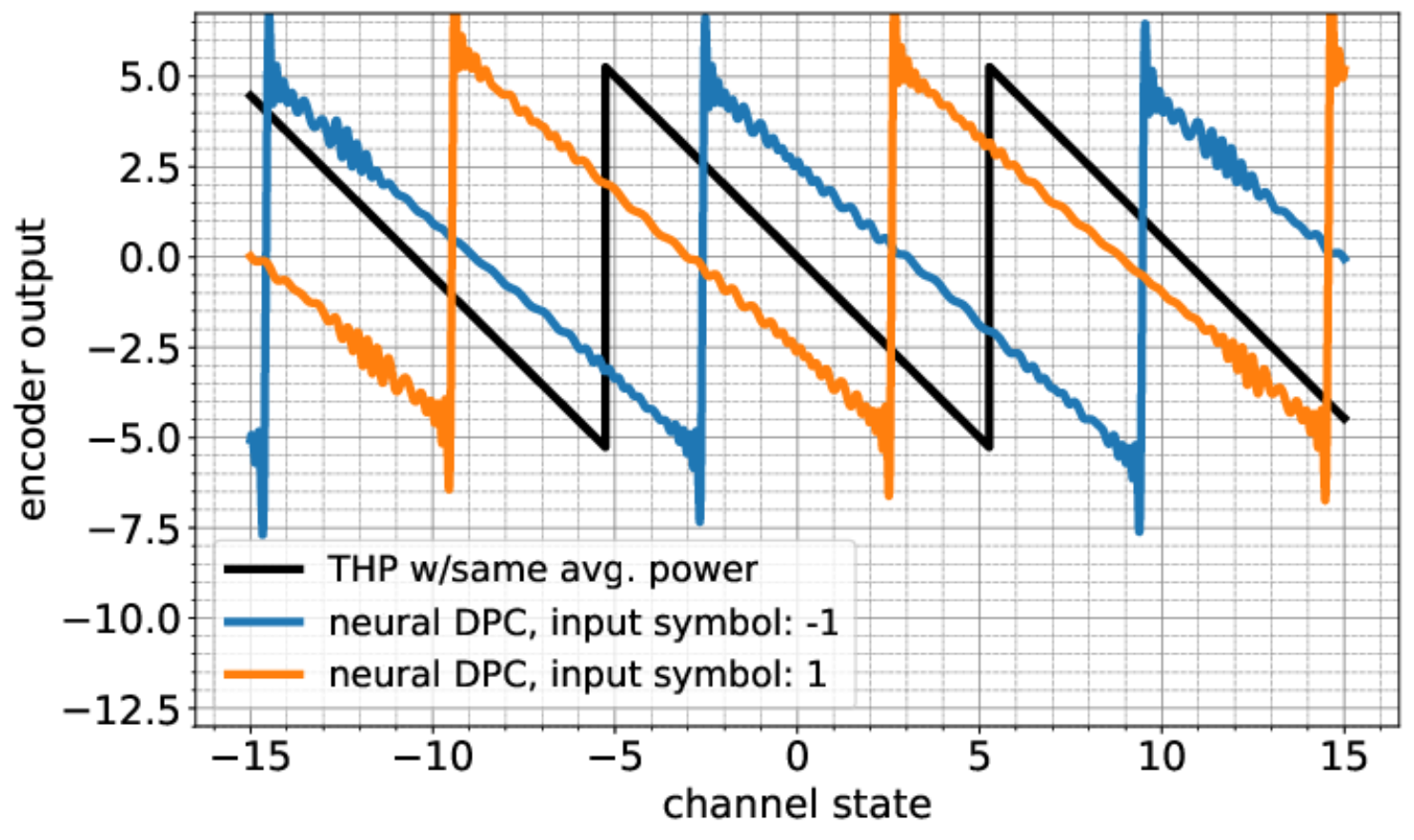}
        \caption{$\lambda=100$, scoring $\mathrm{SNR}:$ 9.65 dB,  $\mathrm{SER}:$ -2.00 dB}
        \label{fig:bpsk_viz_large_snr}
    \end{subfigure}
    \caption{\footnotesize{Visualization of encoder outputs as a function of the channel state realization, where the message $V$ is initially mapped to a BPSK modulation (as in results provided in Fig.~\ref{fig:plot_bpsk}). During training, the interference and channel noise are set as $S \sim \mathcal{N}(0, 30)$ and $N \sim \mathcal{N}(0, 1)$, respectively. The two panels show models trained with different values of the trade-off parameter $\lambda$ in Eq.~\eqref{eq:loss_fn}, resulting  in different power levels $P_X$, corresponding to lower SNR (top) and higher SNR (bottom). In both cases, the learned encoder exhibits a quasi-periodic triangular structure, reminiscent of the modulo operation in THP. As seen in both panels, the encoder's absolute amplitude and periodicity adjust with $\lambda$, illustrating the model’s ability to adapt its encoding strategy across varying SNR.}
}
    \label{fig:encoder_viz_bpsk}
\end{figure}

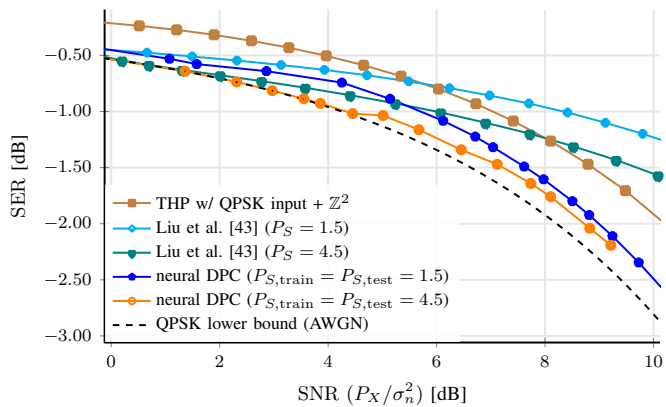
\begin{figure}
\begin{tikzpicture}
\begin{axis}[
	scale only axis,
	font=\normalsize,  
	width=\linewidth,
	height=0.6\linewidth,
	xmajorgrids=true,
	ymajorgrids=true,
	xmin=0,
	xmax=10,
        xtick={0,2,...,10},
	ymin=-3,
	ymax=-0.15,
        y tick label style={
        /pgf/number format/.cd,
            fixed,
            fixed zerofill,
            precision=2,
        /tikz/.cd
    },
    ytick={-3,-2.5, ..., 0},
	xlabel={$\mathrm{SNR}$ $(P_X/\sigma_n^2)$ [dB]},
        ylabel={$\mathrm{SER}$ [dB]},
	legend columns=1,
	legend cell align={left},
	legend style={at={(0,0)}, anchor = south west},
	]

\addplot[line width=2pt, brown, mark=square*] table[x=X, y=Y, col sep=tab] 
    {figures/bounds/QPSK_THP.txt};

\addplot[line width=2pt,  cyan, mark=halfsquare*] table[x=X, y=Y, col sep=tab]{figures/bounds/Structured_QPSK_SER_LIU.txt};

\addplot[line width=2pt,  teal,mark= heart] table[x=X, y=Y, col sep=tab]{figures/bounds/Structured_QPSK_SER_LIU2.txt};

\addplot[line width=2pt,  blue, mark=pentagon*] table[x=X, y=Y, col sep=tab] 
    {figures/data/QPSK_Ours_other.tex};

\addplot[line width=2pt, orange, mark=o] table[x=X, y=Y, col sep=tab]
    {figures/data/QPSK_Ours_other2.tex};

\addplot[line width=2pt, mark=none, black, dashed] table[x=X, y=Y, col sep=tab] 
    {figures/bounds/QPSK_Lower.txt};

\legend{
    THP w/ QPSK input + $\mathbb{Z}^2$,
    Liu et al.~\cite{Liu} ($P_S  = 1.5$), 
    Liu et al.~\cite{Liu} ($P_S = 4.5$), 
    neural DPC ($P_{S, \mathrm{train}} = P_{S, \mathrm{test}} = 1.5$),
    neural DPC ($P_{S, \mathrm{train}} = P_{S, \mathrm{test}} = 4.5$),
    QPSK lower bound (AWGN)
}

\end{axis}

\end{tikzpicture}
\caption{\footnotesize{Symbol error rate (SER) as a function of signal-to-noise ratio (SNR), where both the message $V$ and the interference $S$ are drawn from a fixed QPSK constellation. During training, the interference power is selected from $P_S \in \{1.5,\ 4.5\}$, and the channel noise is modeled as $N \sim \mathcal{N}(0,\ 1.0)$. For the neural DPC model, SNR is varied by training separate models for different $\lambda$ values in Eq.~\eqref{eq:loss_fn}.}}
\label{fig:empirical_results_structured}
\end{figure}

\subsection{Numerical results}

First, we discuss the results based on the Gaussian setting where $S \sim \mathcal{N}(0, \sigma_{s}^2)$. In both panels of Fig.~\ref{fig:empirical_results}, we observe that the proposed neural DPC scheme consistently outperforms THP combined with several aforementioned lattice-based precoding benchmarks, particularly in the low SNR regime. As SNR increases, the performance of neural DPC gradually converges to that of THP. Although the model is trained with a fixed interference variance $\sigma^2_{s, \mathrm{train}}$, it generalizes well across a broad range of unseen test-time conditions (see train/test mismatch curves). As $\sigma_{s,\mathrm{test}}^2$ decreases, the SER gracefully approaches the interference-free optimum, demonstrating the robustness of the learned scheme to varying interference levels.

To better understand the source of these gains, Fig.~\ref{fig:encoder_viz_bpsk} visualizes the learned encoder mappings where the message $V$ is initially mapped to a BPSK modulation. At low SNR (Fig.~\ref{fig:bpsk_viz_small_snr}), the encoder adopts a quasi-periodic triangular structure with independently adjustable amplitude and periodicity, offering greater flexibility than scalar THP, which uses a fixed slope whose amplitude is dictated by the power constraint. This added expressiveness allows the neural encoder to more effectively account for the interference structure, as reflected in the performance gains shown in Fig.~\ref{fig:plot_bpsk}. At high SNR (Fig.~\ref{fig:bpsk_viz_large_snr}), the learned mapping closely resembles that of THP, exhibiting a similar (though shifted) shape, which is consistent with the SER performance convergence observed in Fig.~\ref{fig:plot_bpsk}. The corresponding decoder mapping in this regime (not shown) behaves like a repeated sign function, partitioning the real line into evenly spaced decision regions.

We now examine a Gaussian interference where the message~$V$ is initially mapped to a QPSK modulation. The encoder (not shown) learns a parallelogram-like mapping centered at the origin for each message. The decision regions at the decoder exhibit a near-hexagonal tiling as seen in Fig.~\ref{fig:decoder_viz_sin}. These observations suggest that our neural DPC scheme implicitly recovers a modulo-lattice operation at both the transmitter and demodulator. Unlike prior lattice-based schemes that rely on predefined tessellations (see Sec.~\ref{sec:THP_and_lattice}), the learned model offers greater flexibility, as the densest lattices still remain unknown in most dimensions~\cite{conway1988sphere, zamir, cohn2017sphere} and may vary across operating regimes. Similar to its one-dimensional counterpart (Fig.~\ref{fig:plot_bpsk}), the learned model also remains robust under varying interference conditions, as shown in Fig.~\ref{fig:plot_qpsk}.

In the structured interference setting, where both the message $V$ and the interference $S$ are drawn from a fixed QPSK constellation, the neural DPC model outperforms THP and the scheme proposed in~\cite{Liu} across a wide range of SNRs as shown in Fig.~\ref{fig:empirical_results_structured}. The method of Liu et. al.~\cite{Liu} assumes a fixed $P_S/P_X$ ratio and performs well when this ratio is large; however, its performance degrades outside this regime. On the other hand, THP applies a fixed modulo operation without leveraging the structure of the interference, resulting in poor performance in the low SNR regime. The neural DPC model, however, adapts effectively to varying SNR and $P_S / P_X$ ratios.

\section{Conclusion}
\label{sec:conclusion}

We introduced a learning-based DPC framework that achieves strong empirical SER performance compared to classical precoding methods across a range of interference scenarios. Our results show that the proposed neural DPC models recover key characteristics of THP and lattice-based schemes while surpassing their performance, particularly at low SNR. These findings highlight the potential of incorporating learning to advance practical DPC implementations, including MIMO Gaussian broadcast channels where DPC is essential for achieving sum capacity~\cite{Caire, Yu, Vishwanath, Viswanath}. While we presented initial results under structured interference, future work will explore how learning enables user-specific adaptation, which can be combined with superposition coding~\cite[Chapter 5]{network_info_theo} to improve throughput in multi-user communication settings.

\bibliographystyle{./bibliography/IEEEtran}
\bibliography{./bibliography/references.bib}

\end{NoHyper}
\end{document}